# A Novel Feature Map Enhancement Technique Integrating Residual CNN and Transformer for Alzheimer's Disease Diagnosis

Saddam Hussain Khan[1]*

[1]Artificial Intelligence Lab, Department of Computer Systems Engineering, University of Engineering and Applied Sciences (UEAS), Swat 19060, Pakistan

Email: saddamhkhan@ueas.edu.pk

## A B S T R A C T

Alzheimer's disease (AD) involves cognitive decline and abnormal brain protein accumulation. AD is a significant cause of dementia and necessitates timely diagnosis for effective intervention and treatment. Therefore, CAD systems leveraging deep learning advancements have demonstrated success in disease detection but pose model computational intricacies and the dataset minor contrast, structural, and texture variations challenges. In this regard, a novel hybrid FME-Residual-HSCMT technique is introduced, comprised of residual Convolutional Neural Network (CNN) and Transformer components to capture global and local fine-grained AD analysis using MRI. This approach integrates three distinct concepts: a novel CNN Meet Transformer (HSCMT), customized residual learning CNN, and a new Feature Map Enhancement (FME) strategy to learn diverse morphological, contrast, and texture variations of ADs. The proposed HSCMT at the initial stage utilizes stem convolution blocks stacked with CMT blocks followed by systematic homogenous and structural (HS) operations. The customized CMT block encapsulates each element with global contextual interactions through multi-head attention and facilitates computational efficiency through the lightweight mechanism. Moreover, new inverse residual and stem CNN in the proposed CMT enable effective extraction of local texture information and handling vanishing gradients. Furthermore, in the FME strategy, residual CNN blocks utilize TL-based generated auxiliary and are combined with the proposed HSCMT channels at the target level to achieve diverse enriched feature space. Finally, diverse enhanced channels are fed into a novel spatial attention block for optimal pixel selection to reduce redundancy and discriminate minor contrast and texture inter-class variation. The proposed FME-Residual-HSCMT achieves an F1-score (98.55%), an accuracy of 98.42% and a sensitivity of 98.50%, a precision of 98.60% on the standard Kaggle dataset, and demonstrates outperformance compared to existing ViTs and CNNs methods. The optimal results illustrate the effectiveness of our integrated CNN and transformer framework in learning diverse features and achieving good performance in AD diagnosis.

## 1. Introduction

Alzheimer's disease (AD), a predominant neurodegenerative disorder causing dementia, necessitates early intervention for effective diagnosis [1]. The anticipated rise in global annual diagnoses is forecasted to increase from 45 to 131.5 million by the year 2050 [2]. AD's impact on cognitive function, memory, and irreversible neuron damage is profound [3]. As AD progresses, individuals experience profound behavioral, mood, and personality alterations, eventually leading to a loss of communication and social engagement [3]. These challenges burden patients and their families emotionally, financially and also diminish their overall quality of life.

AD progresses through stages: Early/Preclinical, with possibly absent or mild symptoms; Mild Cognitive Impairment (MCI), is characterized by noticeable memory difficulties but intact daily functioning. In Mild Dementia (MildD), symptoms include memory lapses, task difficulties, communication challenges, and mood swings [4]. Moderate Dementia (MD) results in recognition issues, memory loss, and personality changes, while Severe Dementia leads to complete reliance on caregivers, accompanied by communication and recognition skills loss. Therefore, early diagnosis is pivotal for delaying cognitive decline and enhancing patient quality of life, however, a lack of proven proper treatments [5].

Several invasive or non-invasive screening techniques are employed for the clinical evaluation of AD and monitoring [6]. MRI is favored as a less invasive scanning method and offers comprehensive insights into AD progression [7]. However, MRI manual assessment is difficult due to the complexity and distortion caused by inherent noise during image acquisition making it challenging and time for radiologists to classify AD accurately [8]–[10]. Therefore, computer-aided diagnostic (CAD) methods, leveraging advanced algorithms on diverse medical data, offer an efficient avenue for automatic disease diagnosis and potentially reduce dependence on manual diagnosis by medical professionals [11]–[13].

The rise of AD underscores the necessity for efficient DL-based CAD solutions for automating AD diagnosis from MRI data. Moreover, DL methods, particularly convolutional neural networks (CNNs) and Vision Transformers (ViT), automatically capture latent structural features that offer a data-driven alternative to feature selection methods [14], [15]. The utilization of DL and its proficiency in effectively analyzing images across diverse challenges has captured the interest of researchers aiming to enhance diagnostic systems [16]. ViT, distinct from traditional CNNs, employs self-attention mechanisms without convolution, focusing on crucial regions. Some methods reduce self-attention complexity through local or simplified attention mechanisms [17].

Challenges arise, particularly in medical imaging where available datasets may be limited compared to the extracted features, leading to the curse of dimensionality [18]–[20]. Therefore, the transfer learning (TL) concept has been utilized to reduce the risk of overfitting intensifies in small datasets and impacting model performance on unseen data [21]. Moreover, the datasets face challenges of limited data, infection position, and contrast variation, morphology variability, and inter-class variation and homogeneity with other ADs. Furthermore, conventional ViT methods face challenges in capturing features in the local feature extraction and are computationally complex. Addressing these challenges, this paper introduces FME-Residual-HSCMT, a streamlined DL network based on CNN and Transformer. FME-Residual-HSCMT integrates an early-stage CNN block and a two-stream network design and optimizes information extraction for subsequent transformer processing. The proposed hybrid CNN-Transformer is based on complementary residual learning captured local and global information, and texture variation, respectively [22], [23]. This research study primarily contributes to:

- The novel hybrid technique, named "FME-Residual-HSCMT," synergizes transformer and CNN capabilities to effectively analyze AD in MRI images at both global and local levels. The FME-Residual-HSCMT comprised three heterogeneous concepts: HSCMT, residual learning CNN, and Feature Map Enhancement (FME).

- The proposed HSCMT architecture includes an abstract stem CNN and customized CMT blocks, followed by a progressive arrangement of CNN layers, incorporating homogeneous and structural (HS) operations. Moreover, the multi-head attention concept in the customized CMT computes the implication by encapsulating each entity with global contextual information and capturing their interactions. Furthermore, the light attention operation has been introduced for computational efficiency and allows for efficient local information extraction.

- In the proposed FME-Residual-HSCMT strategy, the novel HSCMT channels are combined with TL-based Residual CNN blocks generated auxiliary to achieve diverse enriched feature space and provide to attention mechanism at the target level. Residual Learning-based CNN blocks with convolutional layers capture local features crucial for distinguishing contrast and texture variations in AD MRIs. Moreover, the inverse residual block concept has been implemented in customized CMT blocks to facilitate gradient propagation in the deeper architecture.

- The novel spatial attention facilitates optimal pixel selection to enhance the subtle discriminative patterns and contrast minor inter-class variations shared among different ADs. Finally, the proposed FME-Residual-HSCMT techniques are rigorously implemented on publicly available Kaggle datasets for effective AD diagnosis and are compared to the best state-of-the-art CNNs and ViTs.

The following sections of the manuscript explore different facets. Section 2 discusses related work, while Section 3 focuses on the proposed AD diagnosis framework. Section 4 provides details on the dataset, implementation, and performance metrics. Section 5 analyzes results and discussions in detail, ablation study, and systematic comparisons. Finally, Section 6 summarizes the manuscript's conclusion and outlines future directions.

## 2. Related work

This section provides a succinct overview of recent automatic AD diagnosis methods utilizing MRI and DL [24]. These existing approaches are trained using either complete or partially cropped MRI images, whereas some methods concentrate on patterns. These methods identify effective patterns on prior knowledge or identification algorithms applied to entire MRI datasets for diagnostic challenges. The DL models are typically initially trained on large-scale natural image datasets like ImageNet and fine-tuned on dedicated medical datasets [25]–[28].

Recent studies lean towards patch-level and subject-level images using DL algorithms for feature extraction and richer information using MRI. Kang et al. [29] employed unsupervised training with DCGAN on coronal-plane 2D MRI scans, achieving 90.36% accuracy and an AUC of 0.897 in the AD challenges. Zhu et al. [30] analyzed MRI into non-overlapping patches, selecting larger t-values for AD classification, and achieving 92.4% accuracy on ADNI. Lian et al. [31] employed an FCN to produce disease risk maps and select high-risk with classification accuracies of 91.90% and 89.80% on ADNI and AIBL, respectively. Hedayati et al. [32] utilized an autoencoder for feature extraction and achieved 95% accuracy on ADNI using another CNN for classification. A study by et al. [33] introduced "DEMNET," achieving a multi-classification accuracy of 95.23% for AD diagnosis using MRI scans. In contrast, Loddo et al. [34] proposed a fully automated model integrating ensemble DL techniques, including AlexNet, ResNet 101, and InceptionResNetV2. Hazarika et al. [35] developed a DL-based method for AD classification with 95.34% accuracy, 96% sensitivity, and 94.67% specificity, highlighting the need for further validation of diverse datasets to ensure generalization and robustness for early AD detection and treatment enhancement. Moreover, Ebrahim integrated deep CNNs AlexNet and ResNet-18 for feature extraction and employed SVM for AD diagnosis, exhibiting 95.10% accuracy and 95.25% sensitivity [36]. Another study utilized deep CNN and MLP for diagnosing AD, leveraging a dataset sourced from Kaggle, and attained an accuracy of 94.61% [37]. HTLML two phases method is implemented that combines TL-based DenseNet for feature extraction and ML voting classifiers, resulting in an accuracy of 91.75%, and an F1-score of 90.25% on the Kaggle AD dataset [38]. Hu et al. [39] introduced the VGG-TSwinformer, a DL model for short-term longitudinal studies utilizing brain MRI as the primary biomarker, predicting MCI progression with 77.20% accuracy. However, recent efforts have focused on training end-to-end DL models and leading to the development of large models that demand high-performance resources. Additionally, traditional DL methods encounter difficulties in capturing local features and navigating computational complexities.

- CNNs may reduce spatial correlation by prioritizing local information, possibly affecting their effectiveness in handling larger and more complex patterns.

- ViTs partition images into linear patches, which may result in sensitivity to patch size and inadequate representation of local details and low-resolution features.

- The training of DL models has long been hindered by the vanishing gradient issue, where gradients tend to diminish when propagated through deeper successive layers.

## 3. Methodology

The proposed framework implements a novel FME-Residual-HSCMT and existing ViTs/CNNs techniques for MRI-based AD diagnosis. Additionally, data augmentation is utilized as a pre-processing technique to mitigate bias and enhance generalization. The proposed AD detection approach involves three experimental setups: (1) a novel FME-Residual-HSCMT, (2) an assessment of existing ViTs and hybrid CNN-ViT, and (3) an evaluation of existing CNNs. The overall AD diagnosis workflow is illustrated in Figure 1.

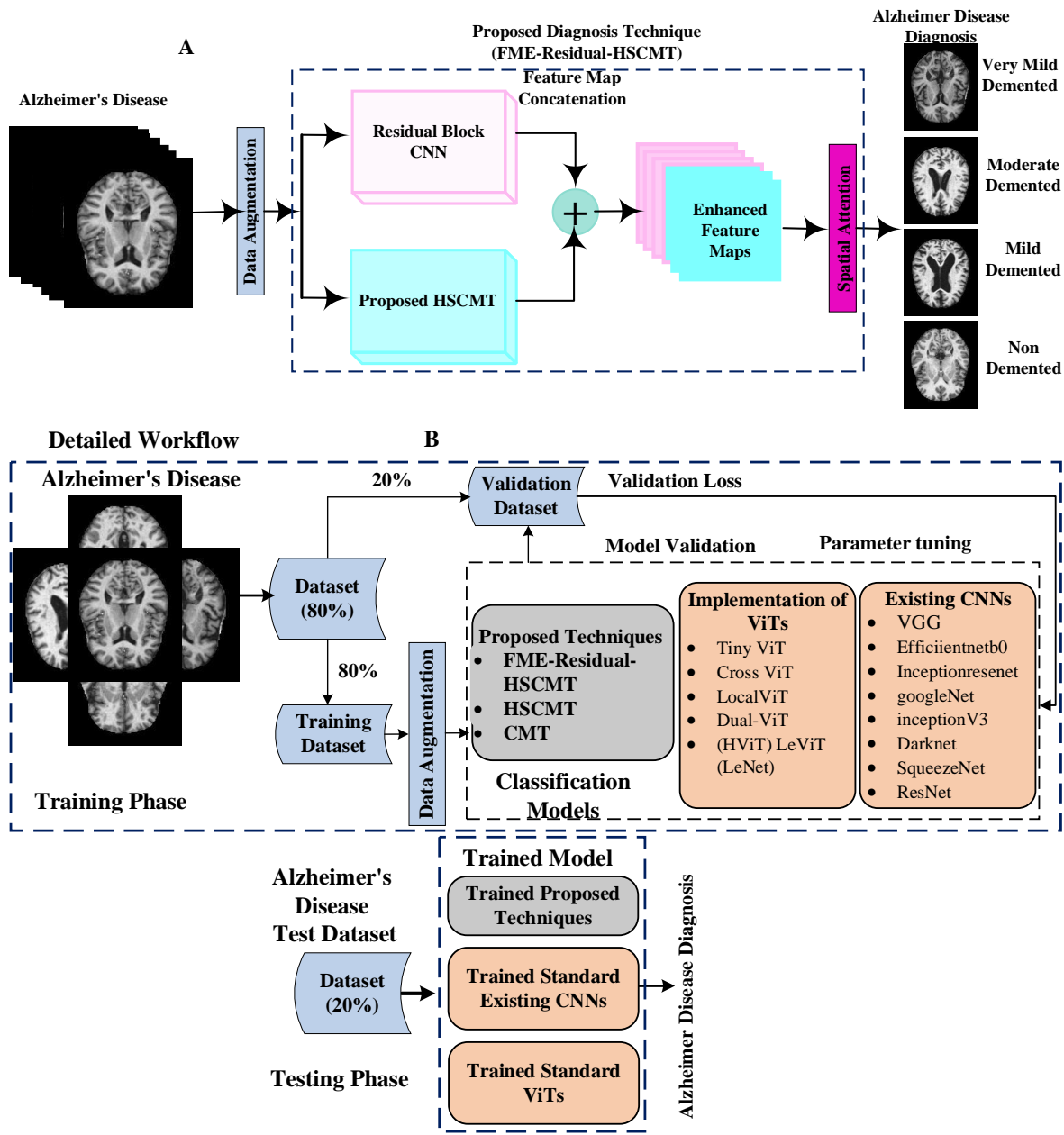

Figure 1: (A) Brief and (B) detailed overview of the proposed AD diagnosis framework.

## 3.1    Data Augmentation

The data augmentation preprocessing phase has been employed on the training data to combat imbalance and overfitting. The dataset poses a challenge due to class imbalance, notably in the MD class, resulting in elevated false positives and affecting outcomes. To rectify this, the data augmentation method is applied to enhance the representation of less contributing classes, effectively mitigating the aforementioned challenges. This encompassed applying random techniques like flipping, scaling, reflection, shearing, etc.

## 3.2    The proposed FME-Residual-HSCMT Technique

The proposed FME-Residual-HSCMT is comprised of new HSCMT architecture and TL CNN Residual blocks for auxiliary feature maps generation and concatenation. The backbone design has four stages, each with a parallel sequence of proposed HSCMT and residual blocks. The proposed HSCMT is inspired by Hybrid ViT-CNN, which has gained prominence in medical image analysis applications. In ViTs, images are segmented into linear non-overlapping patches and processed through encoder blocks using linear layers. However, the linear nature of layers may not effectively capture local and structural information. Therefore, CNNs, with their ability to capture image-specific locality, translational equivariance, and correlated features within two-dimensional neighborhoods, play a crucial role in addressing ViT limitations. In this regard, the new HSCMT is designed specifically to prioritize the efficiency of convolutions, especially in the initial stages of AD image processing for patching and tokenization. Consequently, this emphasis on convolutional efficiency aims to enhance the model's capability to extract relevant features. Moreover, the TL-based residual learning is employed to generate diverse rich, and high-dimensional feature maps. The Feature Map is generated from the proposed HSCMT and residual blocks CNN at the target level and is combined to achieve diverse rich information. Finally, the FME-Residual-CMT Net employs an attention module to focus on subtle minor discriminative patterns and intra-class contrast variation shared among different ADs. The proposed HSCMT and the integration of Residual Learning within the FME-Residual-HSCMT framework are illustrated in Figure 2. Each module's specifics are explained below.

### 3.2.1    The Proposed HSCMT Architecture

Our research endeavors to establish a hybrid network that capitalizes on the synergies of CNNs and transformers. Therefore, the proposed HSCMT architecture integrates an initial stem CNN alongside customized CMT blocks, followed by a sequential deployment of CNN layers integrating both homogeneous and structural (HS) feature extraction processes. The images undergo an initial processing step within the stem CNN block, are divided images into patches, and undergo global feature extraction using embedded patch tokens within the transformer blocks, as depicted in Figure 2. The stem CNN block employs a 3×3 convolution (Conv_L) with a stride of 2 and an output channel of 64. Following this, two additional 3×3 Conv_L with a stride of 1 are applied to further enrich local information extraction.

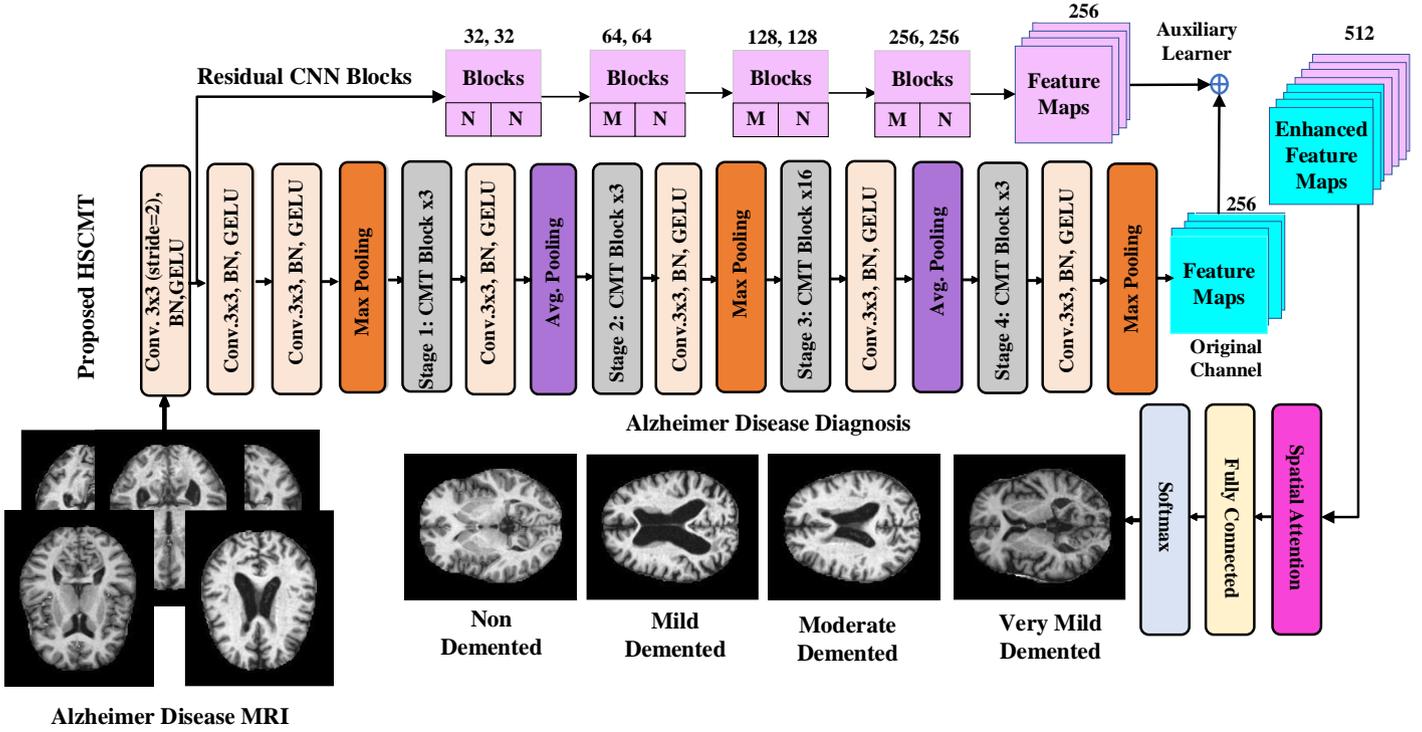

Figure 2: The proposed FME-Residual HSCMT technique, comprised of the proposed HSCMT integrating stem-CNN and customized CMT followed by average and max-pooling, and TL-based residual M and N CNN blocks.

The customized HSCMT architecture is designed with four stages, aiming to produce feature maps of varying scales, particularly pertinent for tasks involving dense prediction. Moreover, Conv_L and activation are utilized after each stage, followed by both max and average pooling, to improve region homogeneity and capture structural information (Equations 1-3). The input image's feature dimension increased to 64 in the initial convolution block and doubled after each pooling operation. This process facilitates the downsampling of the intermediate feature size (resolution) at the conclusion of each stage to capture invariance and improve robustness. Following the initial stages, the patch embedding layer generated various dimensional tokens. Finally, the proposed HSCMT generates four hierarchical channels with distinct resolutions, akin to conventional CNNs.

$$x_{k,l} = \sum_{i=1}^{m} \sum_{j=1}^{n} x_{k+i-1,l+n-1} f_{i,j} \qquad (1)$$

$$x^{max}_{k,l} = \max_{i=1,...,w, j=1,...,w} x_{k+i-1,l+j-1} \qquad (2)$$

$$x^{avg}_{k,l} = \frac{1}{w^2} \sum_{i=1}^{w} \sum_{j=1}^{w} x_{k+i-1,l+j-1} \qquad (3)$$

The Conv_L extracted feature map, represented by '$x$' with dimensions 'k x l,' is characterized in Equation (1) by the kernels denoted as '$f$' with a size of 'i x j.' The output spans [1 to k-m+1, l-n+1]. The homogenous and structural operation window size, denoted by '$w$' is applied to the convolved output $(x_{k,l})$ in Equations 2-3.

### 3.2.1.1 Proposed CMT Block

The proposed CMT architecture, consisting of four stages meticulously designed to produce channels of varying scales is essential for addressing prediction challenges. In each stage, CMT blocks are systematically stacked, facilitating feature transformation, and preserving input resolution. Figure 3 visually represents the incorporation CMT block in each stage. Preceding each stage, a patch embedding layer, incorporating 3×3 Conv_L and normalization (Norm) techniques, establishes a hierarchical representation within the proposed CMT. Additionally, the customized CMT enables the generation of multi-scale representations for downstream tasks, with various CMT configurations and following stridden Conv_L to the input that is particularly advantageous for AD analysis.

Noteworthy is the CMT block's capability to adeptly capture both local and long-range dependencies, as expounded in Section 3.2. In alignment with the ViT concept, the proposed CMT block deviates from the traditional MSA, introducing a lightweight (LMSA) block, while the MLP layer is replaced with a new Inverted Residual Feed-Forward Network (IRFFN). These include two layer-normalization sub-layers followed by LMSA and IRFFN, respectively. The architectural refinement is pivotal in enhancing the ability to capture intricate dependencies and enrich its feature representation for improved performance in complex tasks. Furthermore, to augment the network's representation capacity, a Local Perception Unit (LPU) is integrated into the CMT block, as elucidated in the illustrative depiction presented in Figure 3. Finally, the proposed CMT block is formulated with these three components as follows and are mathematically expressed in Equations 4-6.

$$\boldsymbol{y}_i = \text{LPU}(\mathbf{x}_{i-1}) \qquad (4)$$

$$\boldsymbol{z}_i = LMHSA(LN(\boldsymbol{y}_i)) + \boldsymbol{y}_i \qquad (5)$$

$$\boldsymbol{x}_i = IRFFN(LN(\boldsymbol{y}_i)) + \boldsymbol{z}_i \qquad (6)$$

The output features of the i$^{\text{th}}$ block from LPU and LMHSA blocks are represented as $\boldsymbol{y}_i$ and $\boldsymbol{z}_i$. LN is implemented, and multiple CMT blocks are sequentially stacked in each stage to enable efficient feature transformation and aggregation.

**Customized CMT Design**

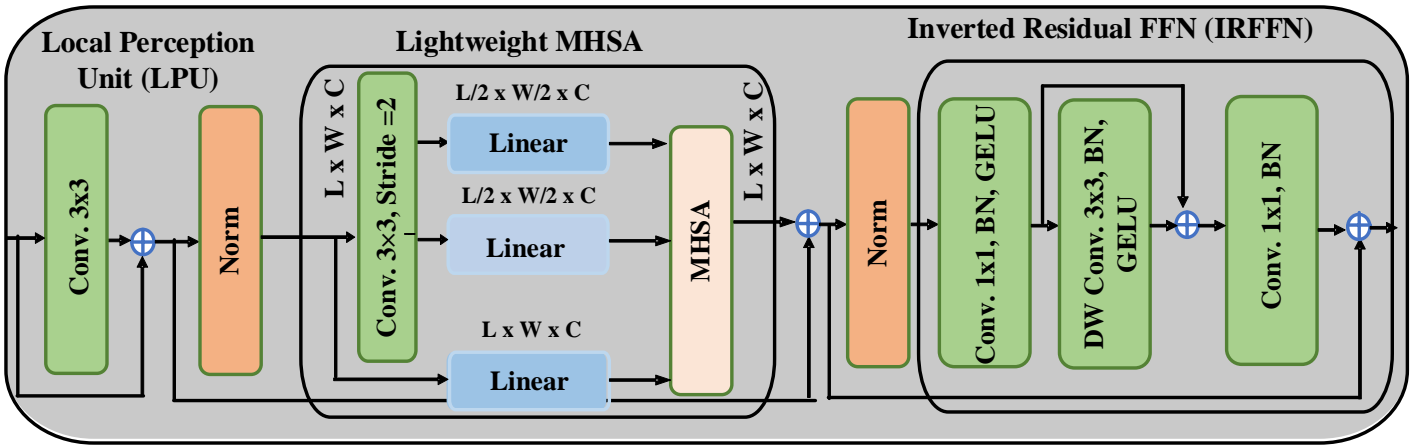

Figure 3: The proposed CMT blocks featuring LPU, lightweight MHSA, and IRFFN.

## A. Local Perception Unit (LPU)

The LPU tackles challenges of invariance, specifically addressing rotation and shift data augmentation and therefore, preserving translation-invariance in model outcomes [27]. In contrast to prior transformers employing absolute positional encoding that introduces unique positional encoding to each patch and disrupts invariance [6]. ViTs, overlooking local relations [38] and structural information [26] within patches, require an improved approach therefore, LPU is introduced to extract local information and overcome these limitations. The input is '**x**' with sizes *L, W, C*. where L×W and d represent the size of the input and feature dimension, respectively (Equation 7).

$$LPU(\boldsymbol{x}) = Conv\_L(\boldsymbol{x}) + \boldsymbol{x} \tag{7}$$

## B. Light Weight Multi-Head Self-Attention (LMHSA)

The MSA enhances the limitations of a single-head Self-Attention (SA) module, which tends to focus on a limited number of positions, potentially neglecting other crucial ones. The SA is the pivotal component of the ViT, renowned for its explicit representation of relationships among entities within a sequence. Numerous ViT architectures have surfaced, modifying the SA module to boost their effectiveness. Certain models incorporate dense global attention mechanisms, while others leverage sparse attention mechanisms to grasp global-level dependencies in spatially uninformative images. Therefore, MSA addresses this constraint by employing parallel stacking of SA blocks to augment the effectiveness of the SA layer. This mechanism computes the implication by encapsulating each entity with global contextual information and capturing their interactions [30]. The initial SA block linearly transforms input map '*x*' to produce query (q), key (k), and value (v) matrices. This involves assigning distinct depiction subspaces (**q**, **k**, and **v**) to the attention layers, allowing MSA to learn diverse and intricate interactions among sequence elements (Equations (8-10)), as detailed in Equation 3.

$$A(\boldsymbol{q}, \boldsymbol{k}, \boldsymbol{v}) = \sigma\left(\frac{q.k^T}{\sqrt{d_k}}\right) \tag{8}$$

$$\boldsymbol{k}' = Conv\_L(\boldsymbol{k}) \tag{9}$$

$$\boldsymbol{v}' = Conv\_L(\boldsymbol{v}) \tag{10}$$

Moreover, a 3× 3 depth-wise Conv_L with a stride of 2 reduces the spatial dimensions of k and v before the attention operation for computational efficiency and allows for efficient local information extraction. The initial stages of the transformer are replaced with convolution blocks, resulting in downsized various-dimensional feature maps. The maps are token-embedded and processed by transformer blocks for feature extraction. A 4 × 4 window patch size is employed in the attention mechanism for enhanced processing granularity. Additionally, a relative position bias **B** is incorporated into the SA block, defining the corresponding lightweight attention (LA) as follows:

$$LA(\boldsymbol{q}, \boldsymbol{k}', \boldsymbol{v}') = \sigma\left(\frac{q.k'^T}{\sqrt{d_k}} + \boldsymbol{B}\right)\boldsymbol{v}' \tag{11}$$

The attention mechanism, denoted by 'LA' and 'σ,' employs activation. Here, **q**, **v**, $\boldsymbol{k}^{T}$' represent query, value,

and transposed key matrices, respectively (Equation 11). Additionally, $\sqrt{d_k}$ serves as a scaling parameter, with '$d_k$' representing the dimension of the key matrix.

The MSA comprises several SA blocks, each incorporating learnable weight matrices for query, key, and value subspaces. Subsequently, in the execution of these blocks, the outcomes are combined and transformed utilizing the trainable parameter $\mathbf{w}^o$, projecting them into the output space. Finally, the proposed CMT exhibits notable adaptability for fine-tuning in various downstream detection challenges. The LMHSA module uses "heads," applying LA functions, producing a sequence of dimension n × d/h. These sequences from multiple heads are then merged into a unified n (resolution/number of patches) × d dimensions of $\boldsymbol{q}, \boldsymbol{k}', \boldsymbol{v}'$ features (Equation 12-13). Where 'h' represents head and 'cat' is a concatenation operation. This process is mathematically depicted as follows:

$$LMHSA(\boldsymbol{q}, \boldsymbol{k}', \boldsymbol{v}') = cat(h_{1,} h_{1,} ..., h_h) . \mathbf{w}^o \qquad (12)$$

$$h_i = LA(\boldsymbol{q}_i, \boldsymbol{k}'_i, \boldsymbol{v}'_i), \text{ where i=1,2, ..., h} \qquad (13)$$

## C. Inverted Residual Feed-Forward Network

The proposed IRFFN is akin to the residual block, optimizing performance by adjusting the shortcut connection's location while incorporating an expansion layer and convolution. Normally, FNN is comprised of two linear layers, separated by GELU activation extracting and integrating complex (Equation 14) [40]. The FFN is used in each encoder block following the SA block. We have introduced 3x3 and point-wise (1x1) Conv_L in the transformer FFN block, reducing parameters and computational complexity while maintaining performance. Moreover, IRFFN introduced a skip connection inverted residual block and replaced the activation function GELU and normalization. The introduction of a shortcut follows the principles of classic residual networks, promoting gradient propagation across layers (Equations 15-16). Additionally, utilizing Conv_L captures an efficient local information extraction. Equation 14 signifies the non-linear activation function GELU, denoted by $\sigma_g$. Here, $\boldsymbol{w}_1$ and $\boldsymbol{w}_2$ represent weights, while b1 and b2 correspond to biases.

$$FFN(\mathbf{x}) = \left( b_2 + \boldsymbol{w}_2 * \sigma_g(b_1 + \boldsymbol{w}_2 * \boldsymbol{x}) \right) \qquad (14)$$

$$IRFFN(\boldsymbol{x}) = Conv(\mathsf{F}(Conv(\boldsymbol{x}))) \qquad (15)$$

$$\mathsf{F}(\boldsymbol{x}) = Conv\_L(\boldsymbol{x}) + \boldsymbol{x} \qquad (16)$$

### 3.2.2 CNN Residual Block

We employed a refined stacking technique that concatenates TL-based CNN residual learning with four M and N blocks for systematic learning enhancement (Figure 4). PWC, utilizing a 1 × 1 kernel size in the M block, facilitates inter-feature map communication and linearly projects Conv_L feature maps into distinct output dimensions. The kernel size of Conv_L is set to 3 × 3 to achieve a local receptive field in both M and N blocks (Equation 1). Concatenating these blocks strategically with CMT in the final stage enables effective exploration of diverse feature spaces. This systematic arrangement, consisting of four sequential residual blocks, facilitates the acquisition of a wide range of essential features. We progressively increase channels from 64 to 256 to fortify the learning process, ensuring a meticulous and refined learning experience for improved outcomes.

The residual block utilizes TL for generating extra feature maps, effectively diversifying channels. These added channels, obtained through TL-based deep CNNs, dynamically capture subtle variations in representation and texture within MRI of ADs. The use of residual blocks enables the discernment of intricate features crucial for distinguishing contrast and texture variations in AD MRIs, facilitated by the impactful deep CNN rooted in FME.

$$\boldsymbol{y} = T(\boldsymbol{x}, \{\mathbf{w}_i\}) + \boldsymbol{x} \qquad (17)$$

$$\boldsymbol{y} = T(\boldsymbol{x}, \{\mathbf{w}_i\}) + \mathbf{w}_s \boldsymbol{x} \qquad (18)$$

The residual block establishes a shortcut connection between the '$\mathbf{x}$' and '$\mathbf{y}$' vectors corresponding to the examined layers (Equation (17-18)). Function $T(\boldsymbol{x}, \{\mathbf{w}_i\})$ denotes the desired residual mapping. In a two-layer model (e.g., as shown in Figure 4), $T = \mathbf{w}_2 \sigma(\mathbf{w}_1 \boldsymbol{x})$, where $\sigma$ is the ReLU activation function. The $T + \boldsymbol{x}$ operation is facilitated by a shortcut connection with element-wise addition, followed by $\sigma(\boldsymbol{y})$. In cases of misalignment, a linear projection denoted by $\mathbf{w}_s$ via the shortcut connections rectifies the discrepancy.

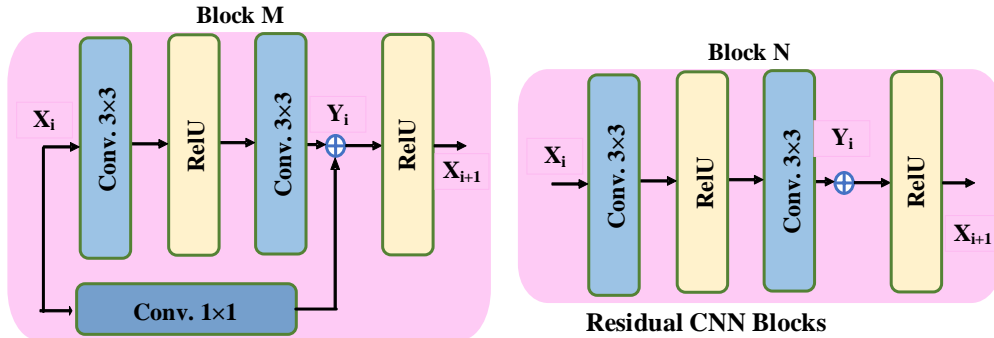

Figure 4: CNN based Residual Block Designing.

### 3.2.3 Feature Map Enhancement

FME is a technique utilized in DL to augment the learning capabilities of the proposed HSCMT technique through TL based residual learning. These auxiliary learners are designed to discern diverse and intricate patterns within images. The proposed FME-Residual-HSCMT Net uses three heterogeneous architectures based on the ideas of HSCMT, residual learning, and FME by concatenating diverse channels (equation 19) to capture multi-level variations. The CNN and CMT based channels specialize in capturing local-level diversity and global target-level features in image patterns, respectively. Moreover, residual learning CNN integrated with FME (FME-HSCMT) captures texture variation and has displayed notable performance for AD analysis. Furthermore, FME-residual learning is employed to retain class-specific information both at the channel and spatial levels. The concepts for channel concatenation are shown in Equation 19, denoted by the symbol ||, and

Figures 1-2.

$$\mathbf{x}_{Boosted} = b(\mathbf{x}_{Residual} \| \mathbf{x}_{HSCMT}) \qquad (19)$$

In Equation (19), the feature-maps of the HSCMT and Residual block are denoted as $x_{CMT}$ and $x_R$, respectively. Additionally, the extra feature-maps from block R, acquired through TL, are represented as $x_E$. The concatenation operation b(.) boosts the feature maps.

### 3.2.4 Spatial Attention-based Approaches

We have utilized a pixel attention (PA)-based block to capture class-specific features at spatial levels (Figure 5). The boosted channels from various learners are amalgamated and weighted through the attention mechanism (Equations 20-22). This allows the network to highlight the most relevant pixels while disregarding those carrying redundant information. Figure, denotes element-wise addition and multiplication, with diverse boosted channel and spatial PA evaluation. The boosted feature map as input is denoted as $\mathbf{x}_{Boosted}$. Refining the boosted feature map involves element-wise addition between various pixel-weighted activations and the input, resulting in computing $W_{pixel}$. Boosted maps are element-wise multiplied with spatial attention, yielding the final refined output feature map $\mathbf{x}_{SA\_out}$. Finally, the sequential weighted output from the SA block is provided with fully connected layers, represented mathematically in Equations (23-24). This process enhances the effective class-specific feature map and discriminative pixel contribution and improves model performance.

$$\mathbf{x}_{SA\_out} = W_{pixel}.\mathbf{x}_{Boosted} \qquad (20)$$

$$x_{relu} = \sigma_1(W_x \mathbf{x}_{Boosted} + W_{SA} SA_{m,n} + b_{SA}) \qquad (21)$$

$$W_{pixel} = \sigma_2(f(x_{relu}) + b_f) \qquad (22)$$

$$\mathbf{x} = \sum_a^A \sum_b^B v_a \, \mathbf{x}_{SA\_out} \qquad (23)$$

$$\sigma(\mathbf{x}) = \frac{e^{x_i}}{\sum_{i=1}^c e^{x_c}} \qquad (24)$$

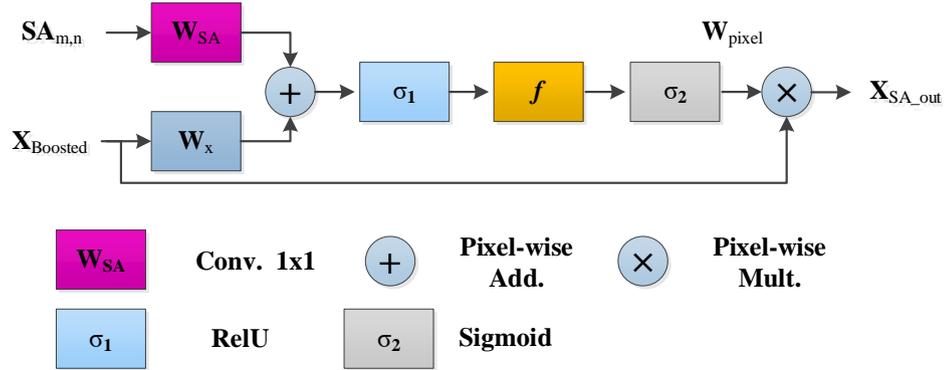

Figure 5: Spatial Attention Block Designing.

In Equation (20), $\mathbf{x}_{Boosted}$ signifies the input map, and $W_{pixel}$ signifies the weighted pixel within the interval [0, 1]. The resulting output, referred to as $\mathbf{x}_{SA\_out}$, highlights the affected region while mitigating the presence of extraneous characteristics. Equations (21) and (22) elaborate on the activation $\sigma_1$ and $\sigma_2$, biases $b_{PA}$ and $b_f$, and transforms $W_x$, $W_{PA}$, $f$. The number of neurons and activation (softmax) in Equation (23-24) is expressed by $v_a$. and $\sigma$.

## 3.3    Implementation of CNNs and ViT Techniques

Our study integrates contemporary ViTs and CNNs models for conducting a comparative analysis. To precisely delineate the ADs in MRI, we employ diverse datasets with various deep CNNs and ViTs [41]. Various CNN and ViT models, including VGG-16/19, ResNet-50, ShuffleNet, Xception, Vit Tiny ViT, Cross ViT, LocalViT, and HVT models among others, are employed for ADs analysis [37]. CNNs have proven effective in detecting AD images in the medical domain [19]. These deep CNNs, featuring diverse depths and network designs, are specifically adapted for analyzing AD types. The ViT models demonstrate heightened accuracy, even on smaller datasets such as CIFAR10. To enhance local feature modeling in ViTs, approaches like LocalViT, and LeViT leverage depthwise convolutions and adaptive position encoding. LeViT (LeNet) incorporates convolutional operations to capture spatial and low-level information.

## 4.    Experimental Setup

## 4.1 Dataset

This research exclusively utilized MRI images authenticated and annotated by professionals and publicly available [41] [42]. Kaggle provides an openly accessible AD dataset with comprehensive information from over 6400 subjects, including MRI scans. This dataset aims to facilitate the development of precise Alzheimer's stage prediction models. In this study, we utilized the ADs [41], featuring MRI images labeled across four classes: MildD, MD, ND, and VMildD. This dataset enables precise training and testing of DL models, enhancing AD stage prediction, and offering the potential for refining diagnosis algorithms and treatment strategies amid the rising global AD burden. Dataset availability on Kaggle sets it apart, stimulating extensive research and algorithm refinement in AD diagnosis and treatment due to its diverse imbalance classes and manageable size. Sample images for each class are shown in Figure 6, with dataset distribution detailed in Table 1. Notably, our study exclusively focused on MRI scans, omitting demographic information and neuropsychological scale scores.

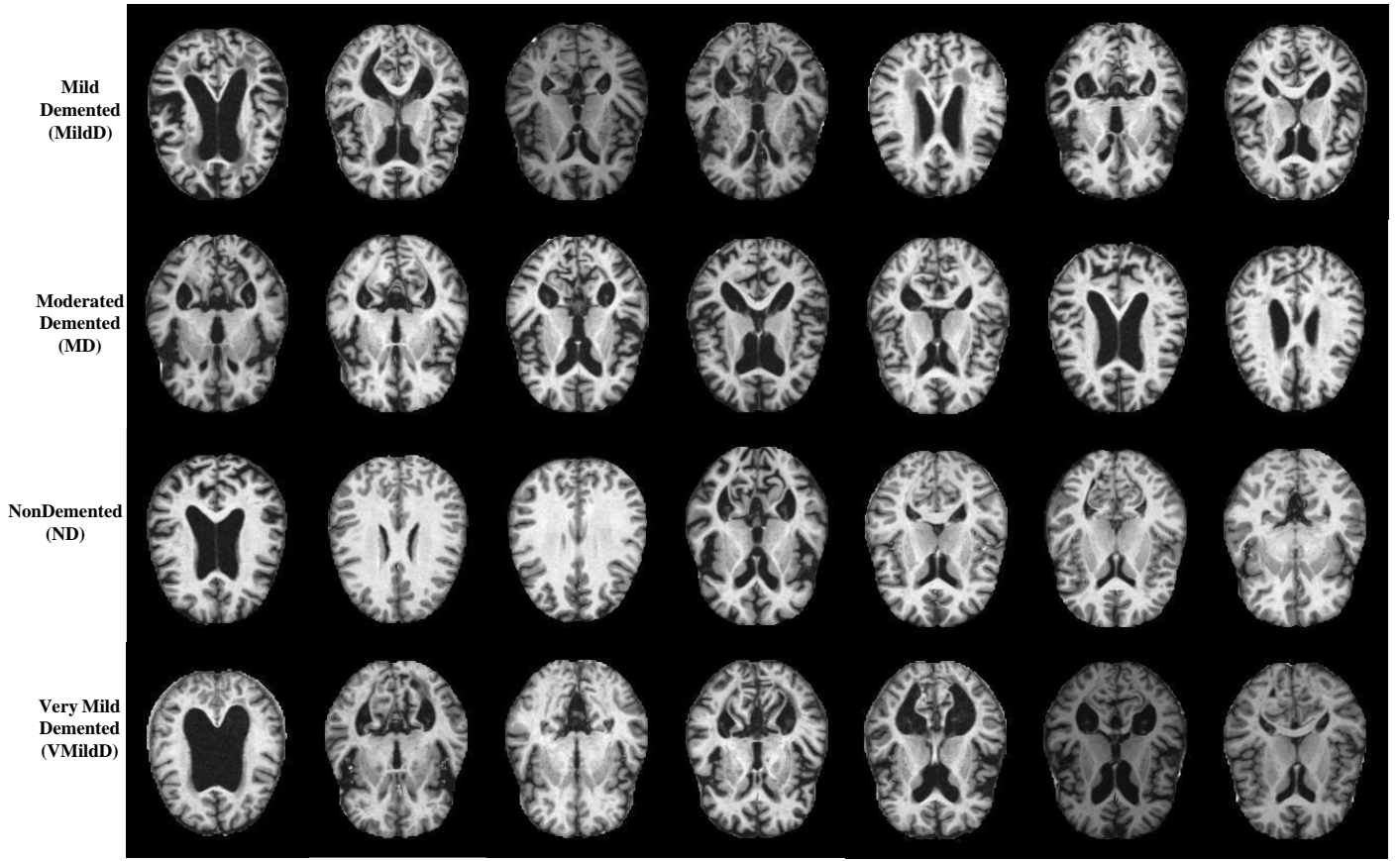

Figure 6: Alzmeir disease labeled MRI images data comprises four classes: MildD, MD, ND, and VMildD.

Table 1. Details of the Kaggle dataset having MRI images of different types of ADs.

| Distributions | MildD | MD | ND | VMildD |
|---|---|---|---|---|
| **Training (80%)** | 574 | 41 | 2048 | 1434 |
| **Validation (10%)** | 143 | 11 | 512 | 358 |
| **Testing (20%)** | 179 | 13 | 640 | 448 |
| **Total (100%)** | 896 | 65 | 3200 | 2240 |

## 4.2 Implementation and Experimental Settings

The proposed FME-residual-HSCMT and existing ViTs/CNNs utilized the Adam optimizer during training, initializing the learning rate at $10^{-3}$ and decay of 85% every 20 epochs while maintaining a constant weight decay of 0.04. To address the class imbalance, the cross-entropy function is employed to evaluate the classification loss, batch size (16), and dropout (0.3) in the output layer. All the experiments have been conducted using Matlab 2023b on hardware featuring an Intel Core i7-10G, 32 GB of RAM, and an NVIDIA GeForce RTX 3060 GPU with 32 GB.

Leveraging data from 6400 subjects in the Kaggle data, performance evaluation on hold-out cross-validation, allocating 20% non-overlapping validation data in each iteration. The results across these validation sets have been aggregated for performance assessment. Models performance has been evaluated using various metrics (Equation 25-28), including F1-score, accuracy (Acc), sensitivity (Sen) /Recall, precision (Pre), ROC/PR curve, and corresponding AUC. Equation (29) delineates the computation of the Standard Error (S.E.) for Sen within a 95% Confidence Interval (CI), aiming to enhance the True Positive (TP) rate and minimize False

Negatives (FNs) in AD's analysis [43]. The z-value of 1.96 represents the S.E. within the 95% CI.

$$\text{Acc} = \frac{\text{TP+TN}}{\text{Total}} \times 100 \qquad (25)$$

$$\text{Sen} = \frac{\text{TP}}{\text{TP+FN}} \times 100 \qquad (26)$$

$$\text{Pre} = \frac{TP}{TP+FP} \times 100 \qquad (27)$$

$$F-\text{score} = 2 \, x \, \frac{\text{Pre} \times \text{Sen}}{\text{Pre} + \text{Sen}} \qquad (28)$$

$$C.I = z \sqrt{\frac{error(1-\text{error})}{\text{Total Samples}}} \qquad (29)$$

## 5. Result and Discussions

This section outlines the evaluation details of the experiments and subsequently, conducted ablation studies to assess the effectiveness of the proposed integrated HSCMT and FME-Residual-HSCMT. Additionally, the performance of FME-Residual-HSCMT is compared to the previous works and existing ViTs/CNNs and two-stream (ViT-CNN) on the benchmarks Kaggle dataset in terms of Acc, Sen, Pre, F1-score, and AUCs, as illustrated in Table 2 and Figures 7-11. The proposed techniques for multi-classification sorts MRI images into four categories: MildD, MD, ND, and VMildD. The proposed HSCMT outperformed existing ViTs/CNNs and achieved an Acc (97.74 %), Sen (96.03 %), Pre (98 %), and F1-score (97). The integration of CNN and transformer in the backbone led to an improvement in performance metrics on the Kaggle dataset. Moreover, the proposed FME-Residual-HSCMT outperformed the customized HSCMT and achieved an Acc (98.42%), Sen (98.50%), Pre (98.60%), and F1-score (98.55%), along with a notable PR-AUC (0.9875) and ROC-AUC (0.9994). The results affirmed the FME effectiveness of the two-stream structure; HSCMT and residual learning, and the early-stage CNN in enhancing feature encoding and generalization performance, as shown in the Figure 7 confusion matrix.

The proposed FME-Residual-HSCMT confusion matrix shows that 636 ND cases are accurately identified. While 2 and 6 MRI images initially labeled as MD and VMildD, respectively, are misclassified as ND. Moreover, 1 and 3 MRI images initially labeled as ND are misclassified as MildD, and VMildD, respectively are correctly classified as MD. Notably, all MD cases are correctly identified, however, 1 label MRI ND is miss-placed. Additionally, 97.20% of MildD cases are accurately classified, with 2.8% misclassified as ND and VMildD, respectively. Similarly, 98.60% of VMildD cases are correctly identified, with 3 MRI images misclassified as MildD and ND, respectively. The overall performance of the proposed techniques in multi-class analysis is summarized in Table 2. Finally, the proposed FME-Residual-HSCMT improves the respective classes detection rate by improving TP while reducing FN and FP of proposed HSCMT and exiting ViT/CNNs.

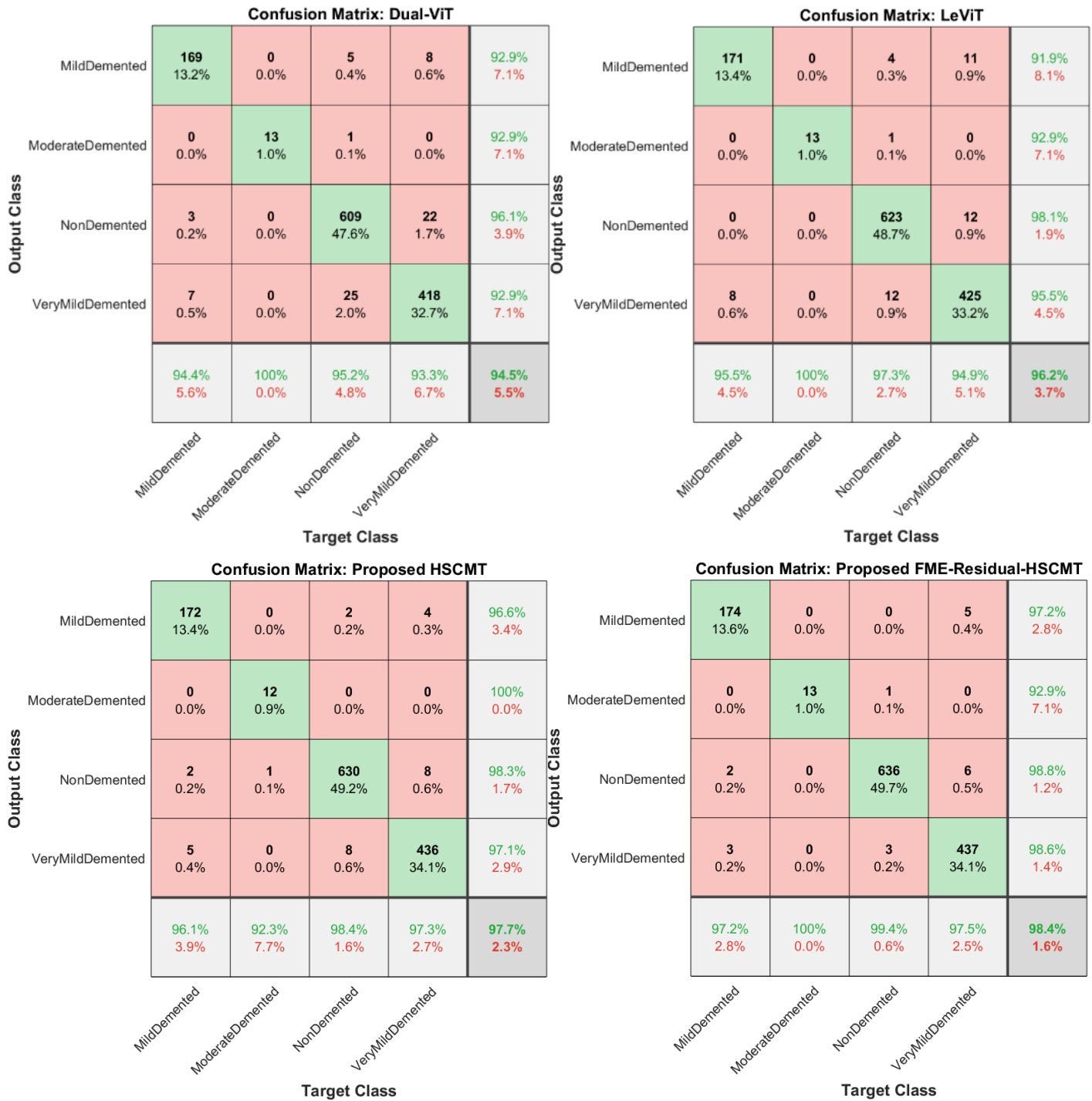

Figure 7: Confusion Matrix of the proposed techniques and existing ViTs/CNNs.

We assessed the complexity of our lightweight and skip connection customized HSCMT model for accurate AD detection using MRI data. These method featuring fewer parameters than traditional CNNs and ViT, with a training complexity, as shown in training plots of proposed FME-Residual-HSCMT, HSCMT, and LeViT in Figure 8. Notably, lightweight HSCMT models required fewer computing resources and training time (approximately 12 minutes per epoch) compared to other ViT models and achieved optimal performance with a streamlined. Moreover, this ensures faster and smoother convergence towards solution and efficient training, suitable for resource-limited hardware, enhancing AD detection accuracy. However, the good performance in the existing hybrid (CNN-ViT) fluctuates during convergence to reach the optimal solution.

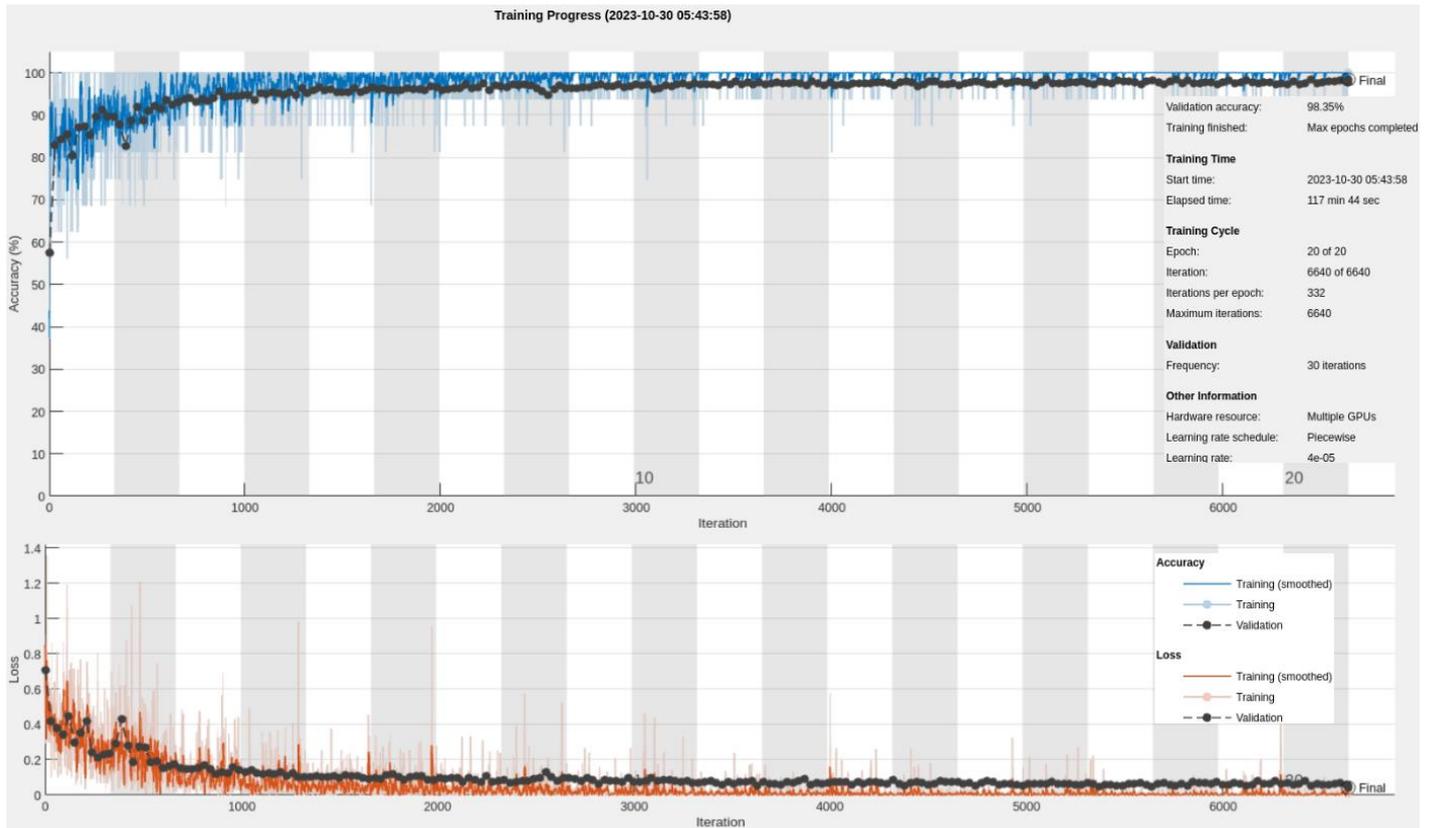

(a)

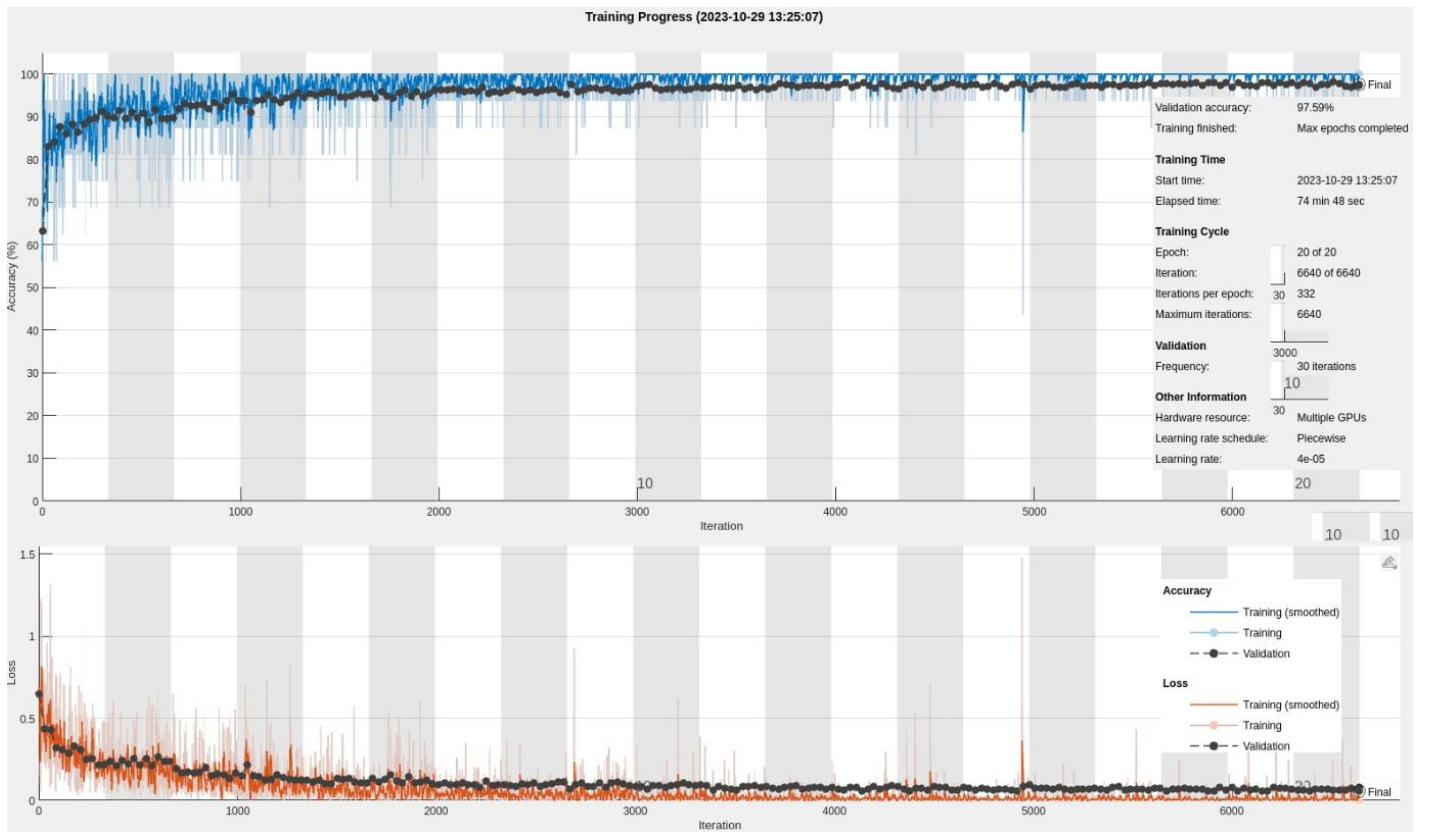

(b)

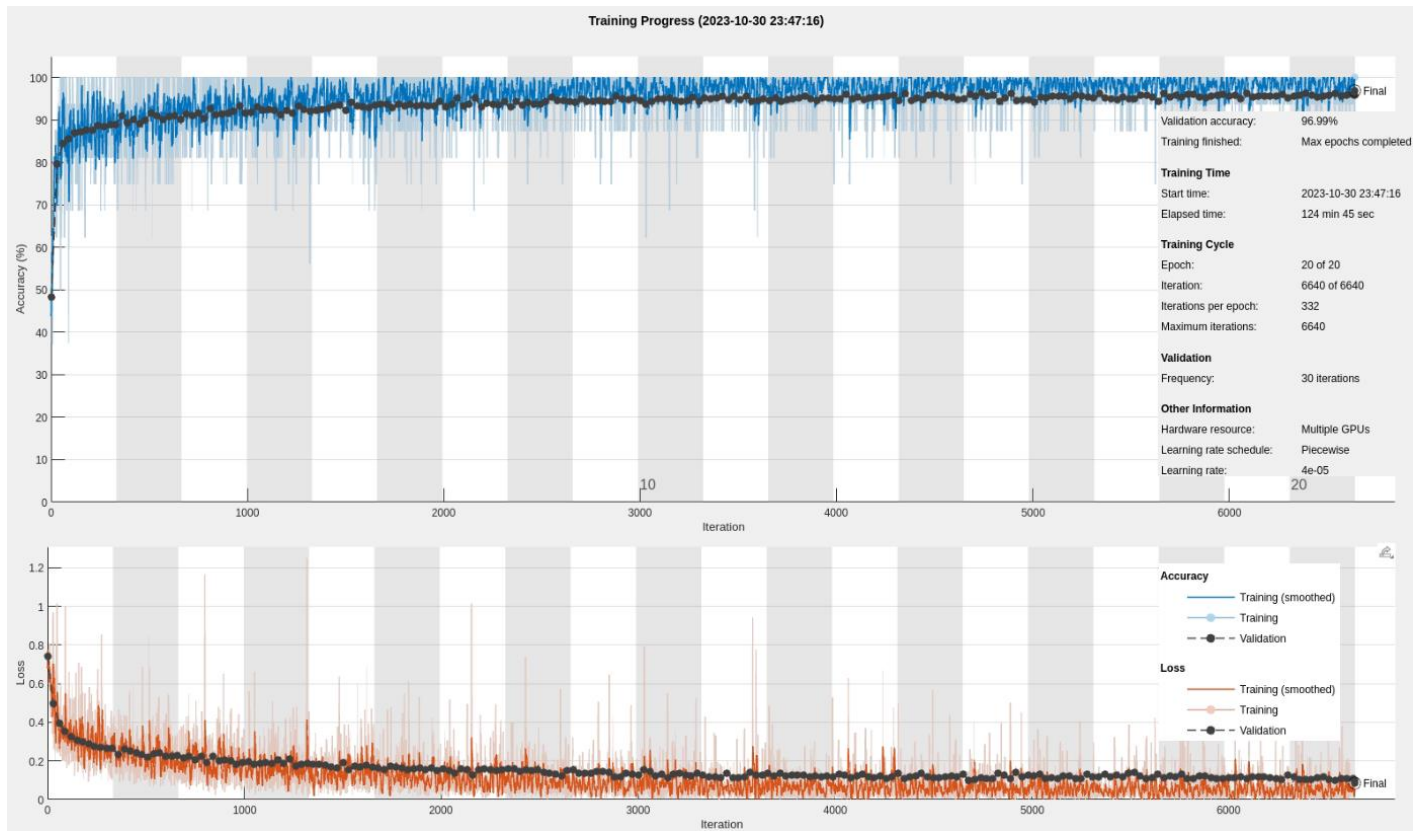

(c)

Figure 8: Training plots; (a) & (b) the proposed FME-Residual-HSCMT & HSCMT, while (c) exiting LeViT.

Table 2. Performance analysis of the proposed techniques and existing CNNs/ViTs.

| Model | Accuracy | Sensitivity | Precision | F1-Score± S.E. |
|---|---|---|---|---|
| **Existing CNNs** | | | | |
| Efficiientnetb0 | 85.55 | 78.66 | 85.57 | 81.97±18.89 |
| Inceptionresenetv2 | 86.95 | 79.85 | 84.58 | 82.14±18.72 |
| GoogleNet | 87.48 | 79.72 | 92.86 | 85.79±15.11 |
| InceptionV3 | 88.59 | 86.2 | 90.34 | 88.22±12.71 |
| Darknet-53 | 88.13 | 87.49 | 90.15 | 88.80±12.13 |
| SqueezeNet | 89.77 | 89.31 | 89.69 | 89.50±11.44 |
| VGG-16 | 91.25 | 90.76 | 89.85 | 90.31±10.64 |
| DarkNet-19 | 93.20 | 94.52 | 94.81 | 94.67±6.33 |
| **Existing ViTs** | | | | |
| Tiny ViT | 89.14 | 88.83 | 90.88 | 89.85±11.10 |
| Cross ViT | 92.97 | 90.28 | 95.59 | 92.86±8.12 |
| LocalViT | 93.08 | 92.83 | 94.59 | 93.70±7.29 |
| Dual-ViT | 94.52 | 95.73 | 93.7 | 94.88±6.12 |
| (Hybrid ViT) LeViT | 96.21 | 96.93 | 94.6 | 95.75±5.26 |
| **Proposed Setup** | | | | |
| **Customized ResNet** | 97.31 | 97.95 | 95.83 | 96.88±4.14 |
| **Proposed CMT** | 97.40 | 97.93 | 97.90 | 97.91±3.12 |
| **Customized HSCMT** | 97.74 | 96.03 | 98 | 97±4.02 |
| **Proposed FME-Residual-HSCMT** | 98.42 | 98.5 | 98.6 | 98.55±2.49 |

Table 3. Performance analysis of the Previous method employed on Kaggle AD's dataset.

| Ref. | Method | Accuracy | Sensitivity | Precision | F1-Score |
|------|--------|----------|-------------|-----------|----------|
| Ahmed et al. (2022) [44] | CNN | 90.00 | 87.34 | 91.34 | 88.09 |
| Sharma et al. (2022) [38] | DL | 91.75 | -- | -- | -- |
| Tanjim .et.al (2024)[45] | VGGs Ensembles | 92 | 88 | 90 | 87 |
| Tanjim .et.al (2024)[45] | DenseNet Ensembles | 95 | 90 | 91 | 89 |
| Balasundaram et al. (2023)[46] | DL models | 94.10 | 94.75 | 96.50 | 95.50 |
| Bangyal et al. (2022)[37] | CNN | 94.63 | 94.75 | 94.75 | 94.50 |
| Mohammed et al. (2021) [47] | DL + ML | 94.80 | 93.00 | -- | -- |
| Murugan et al. (2021) [33] | CNN | 95.23 | 95.00 | 96.00 | 95.27 |
| Ahmad. et. al. (2023) [48] | Lightweight +CNN | 95.93 | 95.88 | 95.93 | 95.90 |
| Tanjim .et.al (2024)[45] | Customized Ensemble | 96 | 93 | 89 | 91 |
| Xing mu. et.al (2023) [49] | CNN+ViT | 97.43 | -- | --- | --- |
| Loddo et al. (2022) [34] | Deep Ensemble | 97.71 | 96.67 | -- | -- |

## 5.1 Performance Comparison with Existing ViTs/CNNs

This section provides a comprehensive comparison of the proposed FME-Residual-HSCMT technique for AD detection with other leading models evaluated on the Kaggle datasets. Various CNN and ViT models, including VGG-16/19, ResNet-50, ShuffleNet, Xception, Tiny ViT, Cross ViT, LocalViT, Dual ViT, and HVT models among others, are employed for AD analysis. Table 2 presents our model's acceptable performance in multi-classification tasks with previous advanced DL methods on the identical dataset. The proposed technique illustrated substantial improvements over conventional local correlated feature learning CNNs, with improvements ranging from 5.22% to 12.87% for Acc, 3.98% to 19.84% for Sen, 3.79% to 14.02% for Pre, and 3.88% to 16.58% for F1-score (Figure 9). Additionally, the proposed technique outperformed existing global receptive learning ViTs with enhancements in the Acc range (5.22% -12.87%), Sen (2.77%-9.67%), Pre (3.01%-7.72%), and F1-score (3.67% to 8.7%), along with a notable increase of 0.62% in PR-AUC. Moreover, a comparative analysis of the most recent previous techniques employed on Kaggle AD's is mentioned in Table 3. Table 2 reveals recent transformer-based Dual-ViT models exhibit relatively larger performance compared to existing CNN models. Existing Dual-ViT also outperforms other ViT models across most metrics, particularly excelling in Acc and F1-score (94.52% and 94.88%), and PR-AUC at 0.9566. In contrast, the LeViT method, rooted in a backbone network based on LeNet CNN and ViT methods, incurs a relatively lower computational overhead while achieving good performance compared to existing CNNs and ViTs.

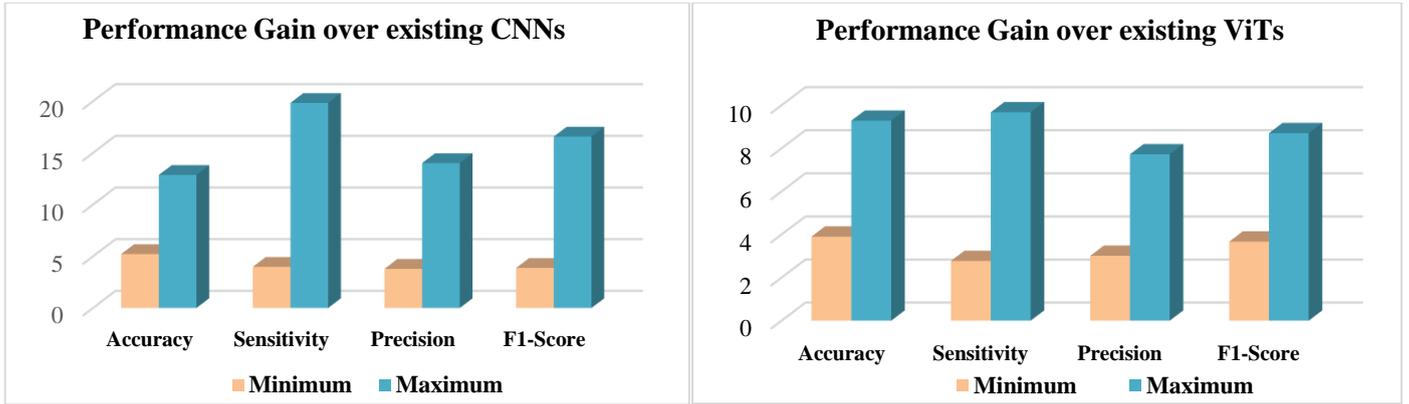

Figure 9: The proposed FME-Residual-HSCMT performance gain over existing techniques.

## 5.2 Ablation Study of the Proposed Technique

In FME-Residual-HSCMT, we optimized convolutional and transformer blocks using novel HSCMT and TL-based residual learning CNN. To assess their impact on performance and efficiency, an ablation study has been conducted with varied block configurations, presenting detailed results in Table 2. Existing CNNs lack global receptive fields, while ViT models lack efficient feature extraction, therefore, the integration of the hybrid LeViT (CNN-ViTs) facilitated the extraction of both local and global learning capacities, thereby further enhancing performance but facing computational complexity. Moreover, the absence of the early-stage CNN method in the backbone resulted in a slight degradation of the model's generalization ability on unseen data. In this regard, we substituted them with standard STEM-CNN blocks and customized lightweight transformer blocks in a hybrid approach. Consequently, the proposed lightweight CMT technique demonstrated notable performance gains over the hybrid CNN-ViTs (LeViT), with improvements in Acc by 2.21%, Sen by 1.57%, Pre by 4%, and F1-score by 2.8%.

Moreover, the incorporation of systematic homogeneous and structural (HS) feature extraction operation with customized CMT using the AD dataset enhances performance from Acc (96.61% to 97.74%), and F1-score (96.06-97%). Consequently, shows improved classification performance, with gains of Acc (3.22-8.6%), Sen(0.3-7.2%), Pre(2.41-7.12), F1-score(2.12-7.17%) over transformer-based models. Finally, the proposed FME-Residual-HSCMT incorporates lightweight techniques, a two-stream structure, and residual learning, to simplify the backbone network and counter overfitting. The boosting of the proposed HSCMT feature-maps with residual learning CNN surpasses all the state-of-the-art ViTs and CNNs methods, evident from performance metrics, ROC-PR curves, and PCA-based analysis (Tables 2-3 and Figures (7-11)).

## 5.3 PR/ROC curves

Detection rate curves quantitatively evaluate the discrimination ability of FME-Residual-HSCMT across various threshold setups, assessing its generalization between moderate and other classes. PR/ROC curves instantly analyze the accurate prediction rate of the AD classes (MildD, MD, VMildD) with ND (Figure 10). PR and ROC curves are plotted by comparing predicted probabilities to ground-truth labels, with method performance summarized in Table 2 at 20% label availability. FME-Residual-HSCMT excels with 0.9875 PR-AUC, and 0.9994 AUC-ROC, surpassing all other active learning methods. Additionally, FME-Residual-HSCMT's AUC-PR improvement ranges from 3% to 11.8%, with AUC-ROC gains of 1.84%. In Figure 10, our FME-Residual-HSCMT excels with the best ROC curves and the highest AUC among the compared CNNs/ViTs methods. Notably, superior ROC performance is indicated by the deep blue curve and points closer to the top-left corner, while better PR performance is represented by the green curve and points nearer to the top-right corner.

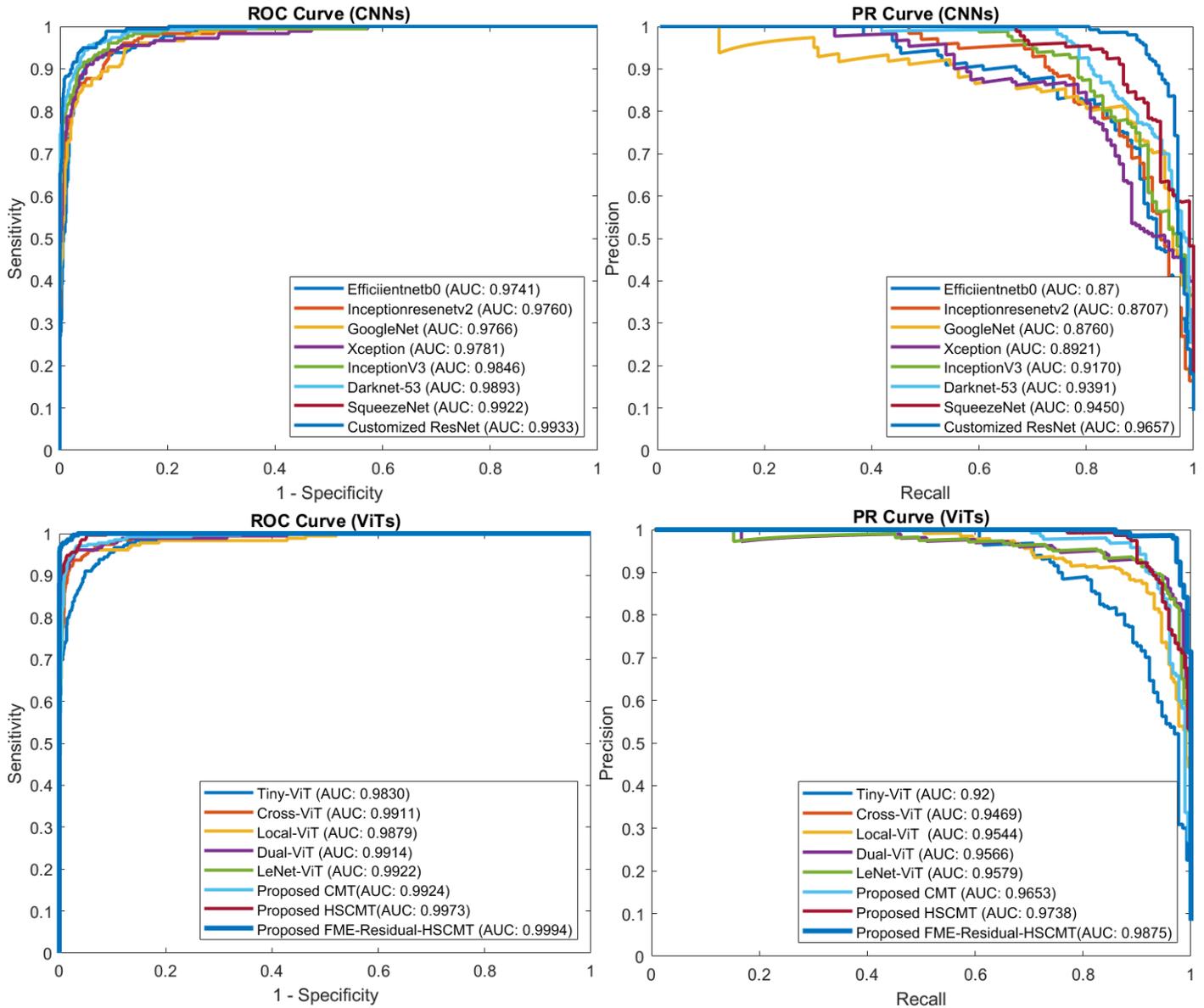

Figure 10: Detection rate analysis of the proposed and existing CNNs/ViTs.

## 5.4 Feature Space Visualization

We extracted feature maps from each layer of the FME-Residual-HSCMT model using the unseen test set to validate the stated argument. However, direct visualization is hindered by the features' high dimensionality. Thus, we employ average pooling and fully connected layers followed by dimensionality reduction PCA techniques to plot them in 2D space. PCA preserves local feature geometry better due to its nonlinear nature, leading to enhanced clustering quality and visualization. The PCA is implemented to visualize embedded features from the second last fully connected layer for four typical methods (FME-Residual-HSCMT, HSCMT, Dual-ViT using the Kaggle datasets in Figure 11. Figure's PCA plots highlight the FME-Residual-HSCMT model's capacity to extract separable features, resulting in clear MildD, ND, MD, and VMildD class separation in the final layers.

This highlights the effectiveness of the hybrid HSCMT-guided contrastive learning scheme in enhancing feature discriminativeness. Figure 11 illustrates PCA embeddings: the proposed HSCMT(a) with and (b) without the FME term, (c) existing Dual-ViT where red, blue, magenta and blue points denote MildD, ND, MD, and VMildD, respectively. Our model's hypergraph embeddings (PC1 Vs PC2) exhibit greater separability than other models (PC1 to PC2), indicating superior discriminative feature representation and enhanced classification performance.

## 5.5 The Proposed Technique's Significance

- The proposed FME-Residual-HSCMT Net leverages three heterogeneous architectures: HSCMT, residual learning, and FME, by concatenating diverse channels to capture multi-level variations.

- Our proposed HSCMT method, inspired integrated approach of CNNs and transformer models, captures global and locally correlated features by stacking multiple CMT blocks. CNN-based channels focus on local-level diversity, while CMT-based maps emphasize global target-level features. Moreover, heterogeneous and structure operations tackle minor contrast and morphological information for inter-class variation and improve the robustness.

- We have introduced convolution blocks and lightweight versions of transformer blocks to enhance efficiency, reducing computational complexity and model capacity. Moreover, LPU preserves invariance to enhance robustness, MSA computes global contextual interactions, and convolutional inverse residual connection captures local texture information in the CMT block.

- TL based FME-residual learning CNN is utilized to preserve class-specific information at both channel and spatial levels and effectively captures texture variation. Moreover, TL and data augmentation address limited and imbalanced data challenges.

- A novel pixel attention-based block captures class-specific features at spatial levels from boosted channels of various learners. The network highlights relevant pixels, disregarding redundant information, enhancing the effective class-specific feature map, and discriminative pixel contribution.

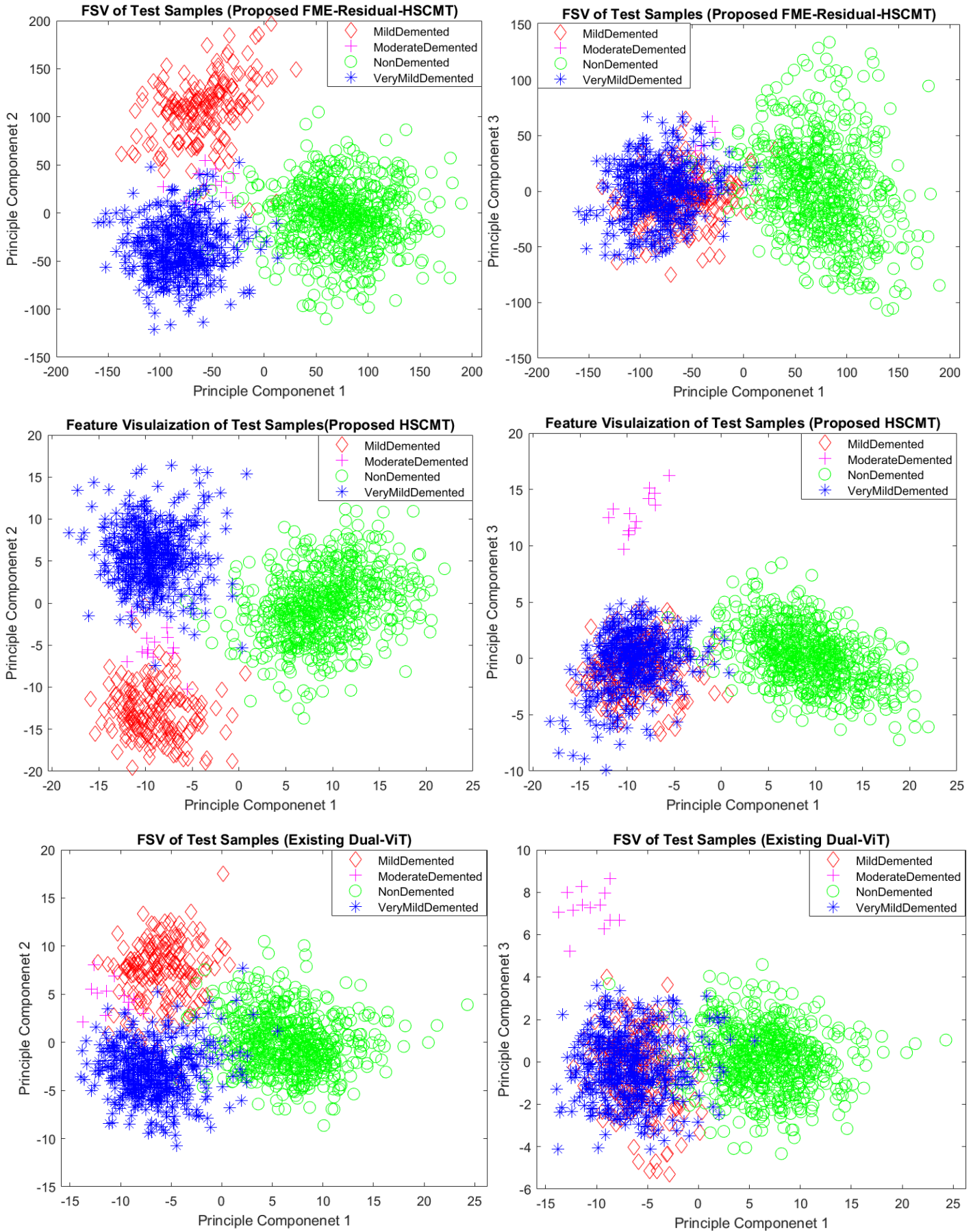

Figure 11: Feature Space Visualization of the proposed techniques on test samples.

# 6. Conclusion

This study presents an efficient integrated STEM and residual CNNs, and a transformer technique for automated AD diagnosis using MRI. The AD data is challenging due to its complex nature, characterized by subtle minor contrasts, and morphological, and texture variations between the classes. Therefore, a three-stream integrated FME-Residual-HSCMT approach CNNs and Transformer method and TL-based residual FME learning are employed to enhance the model's ability to extract diverse features space. Consequently, the proposed technique demonstrated performance gains over the highest performance of existing hybrid LeNet CNN-ViTs, with improvements in Acc by 2.21%, Pre by 4%, and F1-score by 2.8%. However, hybrid CNN and ViT models are computationally complex and often face the challenges of vanishing gradients. Therefore, the proposed HSCMT integrates STEM convolution, lightweight, and transformer blocks accompanied by downsized morphological operations to improve efficiency and reduce computational complexity. The proposed HSCMT architecture is strategically devised with an initial convolutional block for patch embedding and tokenization, and processed by transformer blocks for comprehensive global feature extraction. The lightweight techniques notably reduced computational overhead while preserving model performance and the Residual-FME-RECMT showcased superior efficiency. Additionally, residual learning-based transformers and TL CNNs capture global and local texture variation and mitigate vanishing gradients. Finally, novel spatial attention facilitates optimal pixel selection to enhance the discrimination of minor contrast and patterns of inter-class variation shared among different ADs. In the future, we will incorporate non-imaging data, like gender, age, and neuropsychological scales, to complement MRI data in AD research. Moreover, challenges stem from the limited availability of genetic information, biomarker data, and MRI images, impeding their integration into deep learning methods. Finally, research may emphasize collecting more multi-modal data or devising data imputation algorithms like GANs to enhance model performance.

## Acknowledgment

We thank the Artificial Intelligence Lab, Department of Computer Systems Engineering, University of Engineering and Applied Sciences (UEAS), Swat, for providing the necessary resources.

**Conflicts of interest**: The authors declare that they have no known competing financial interests or personal relationships that could have appeared to influence the work reported in this paper.

## Institutional Review Board Statement

Not applicable.

## Informed Consent Statement

Not applicable.

**Data Availability Statement** Correspondence and requests for materials should be addressed to Saddam Hussain Khan.